\documentclass[aps,prc,onecolumn,showpacs,superscriptaddress]{revtex4}
\usepackage[latin1]{inputenc}
\usepackage{bm}
\usepackage{epsfig}
\usepackage{amsthm}
\usepackage{amsfonts}
\usepackage{float}
\usepackage{amsmath,amssymb}
\usepackage{color}
\usepackage{slashed}
\usepackage{relsize}
\usepackage[toc,page]{appendix} 

\usepackage{graphicx}
\usepackage{dcolumn}
\usepackage{bm}
\newcommand{\be}{\begin{equation}} 
\newcommand{\ee}{\end{equation}} 
\newcommand{\bea}{\begin{eqnarray}} 
\newcommand{\eea}{\end{eqnarray}} 
\newcommand{\nn}{\nonumber} 
\newcommand{\mintedim}[2]{{\int\kern-0.50em\mbox{{\small$\mathop{\frac{\mbox{{\small${\rm d^{#2}}\vect{#1}$}}}{\mbox{{\small$(2\pi)^{#2}$}}}}$}}\ }} 
\newcommand{\inteonedim}[1]{{\int_0^\infty\kern-1em\mbox{{\small${\rm d}{#1}$}}}} 
\newcommand{\intecontour}{\int_{\gamma-i\infty}^{\gamma+i\infty} \frac{ds}{4\pi i}} 
\newcommand{\intet}{{\int_{0}^{\infty}{dt }}} 
\newcommand{\vect}[1]{\bm{#1}} 
\newcommand{\modz}{\frac{1}{1-q}}

\newcommand{\ordertwoarg}{\left(1+\frac{q-1}{q}\frac{\Lambda-\varepsilon_p+3\mu}{T}\right)}

\begin{document}
\title{Analytical Results for the Classical and Quantum Tsallis Hadron Transverse Momentum Spectra: the Zeroth Order Approximation and beyond}

\author{Trambak Bhattacharyya}
\email{bhattacharyya@theor.jinr.ru}
\affiliation{Bogoliubov Laboratory of Theoretical Physics, Joint Institute for Nuclear Research, Dubna, 141980, \\
Moscow Region, Russia}
\author{Alexandru~S.~Parvan}
\email{parvan@theor.jinr.ru, parvan@theory.nipne.ro}
\affiliation{Bogoliubov Laboratory of Theoretical Physics, Joint Institute for Nuclear Research, Dubna, 141980, \\
Moscow Region, Russia}%
\affiliation{Horia Hulubei National Institute of Physics and Nuclear Engineering, Bucharest-Magurele, Romania}

\begin{abstract} 
We derive the analytical expressions for the first and second order terms in the hadronic transverse momentum 
spectra obtained from the Tsallis normalized (Tsallis-1) statistics. We revisit the zeroth order quantum Tsallis distributions and 
obtain the corresponding analytical closed form expressions. It is observed that unlike the classical case, the analytical closed forms
of the zeroth order quantum spectra do not resemble the phenomenological distributions used in the literature after $q\rightarrow q^{-1}$ 
substitution, where $q$ is the Tsallis entropic parameter. However, the factorization approximation increases the extent of similarity.
\end{abstract}
\pacs{12.38.Mh, 12.40.Ee}

\maketitle
%

\section{Introduction}

Hadronic spectra resulting from high energy collision events follow a power-law pattern, 
and the power-law formulae inspired by the Tsallis statistics \cite{Tsal88} are particularly popular while describing the hadronic distributions in the 
phenomenological and experimental studies. 

It is important to understand the origin of the phenomenological Tsallis distributions~\cite{cleymansworku,Bediaga00,Beck00} from the fundamental theories 
and recently there have been some attempts~\cite{Parvan16,Parvan2017a,Parvan19,Parvan20a,tsallistft} to address this issue. In Refs.~\cite{Parvan16,Parvan2017a,Parvan19}, it has been shown that the most generalized form of the hadronic transverse momentum spectra calculated from the Tsallis statistical mechanics is given by an infinite summation and 
the zeroth order truncation of the Maxwell-Boltzmann spectrum yields the widely used expression of the phenomenological Tsallis distribution given, for example, in Refs.~\cite{TsallisTaylor,cleymansworku}. However, in Ref.~\cite{Parvan20a}, it was demonstrated that the phenomenological Tsallis distribution~\cite{cleymansworku,Bediaga00,Beck00} for the Maxwell-Boltzmann spectrum corresponds also to the zeroth term approximation of the $q$-dual statistics based on the $q$-dual entropy obtained from the Tsallis entropy under the multiplicative transformation of the entropic parameter $q \to q^{-1}$.  

Approximating the Tsallis transverse momentum spectra with the zeroth order term is called the `zeroth order approximation', which may work very
well for certain collision energy regions. However, it has been shown~\cite{Parvan16,Parvan19} that the zeroth order approximation may not be sufficient always. In some of the cases, there is a necessity to include the higher order terms in the transverse momentum distribution. 

The present paper calculates the analytical expressions for the first and the second order terms in the Tsallis Maxwell-Boltzmann transverse momentum distribution as they may be indispensable in the phenomenological and experimental studies for describing the data obtained at certain collision energies. While calculating the transverse momentum distributions, we consider the Tsallis normalized (or Tsallis-1) statistics. Tsallis-1 statistics is one of the several schemes in the Tsallis statistical mechanics \cite{Tsal98} which differ in the definition of the average values. 
We discuss this scheme in detail in the next section. 

The paper also calculates the closed analytical form of the zeroth order term of the quantum Tsallis transverse momentum spectra and verifies that unlike the classical case, the quantum Tsallis phenomenological distributions \cite{millerTsFD,TsFDPLA} used in the literature
are not identical with this zeroth order spectra (see also \cite{Parvan19}). However, the factorization approximation of the zeroth order term increases the extent of its similarity with the quantum phenomenological Tsallis-like distributions. 

The remainder of the paper contains a discussion of the basic mathematical set up and the expression for the generalized Tsallis transverse momentum spectra in sections \ref{sec2}, and \ref{sec3}. Analytical calculations of the first and the second order terms in the Maxwell-Boltzmann Tsallis transverse momentum distribution are shown in section \ref{sec4}. Closed analytical forms of the zeroth order Tsallis quantum distributions and their factorization approximation have been discussed in sections \ref{sec5}, and \ref{sec6}. Sections \ref{sec7}, and \ref{sec8} are devoted to the discussions of the results, summary and the outlook.

\section{{Basic definitions and formulae}}
\label{sec2}

The Tsallis statistical mechanics is based on the following definition of entropy~\cite{Tsal88,Tsal98},
\be
S =\mathlarger{ \mathlarger{ \sum}}\limits_{i} \frac{p_{i}^{q}-p_{i}}{1-q},
\label{t1entropy}
\ee
where $q$ is a real parameter, and the probabilities of micro-states $\{p_i\}$ follow the normalization,
\be
\phi=\sum\limits_{i} p_i -1=0.
\label{probnorm}
\ee
The definition of average expectation values in the Tsallis normalized (or the Tsallis-1) scheme 
is given by,
\be \label{t1av}
\langle Q \rangle = \sum\limits_{i} p_i Q_i .
\ee
Here and throughout the paper we use the system of natural units $\hbar=c=k_{B}=1$. When $q\rightarrow1$ the generalized entropy 
given by Eq.~\eqref{t1entropy} reduces to the Boltzmann-Gibbs entropy, $S_{\text{BG}}$=$-\sum_{i} p_i \ln p_i$. Though the parameter 
$q$ may assume values from 0 to $\infty$, yet as far as the description of the hadronic spectra in high energy collisions is concerned, 
we shall be interested in the values of $q<1$ for the Tsallis-1 statistics.

For the grand canonical ensemble, the equilibrium probability distribution $\{p_i\}$ of microstates of the system, its normalization equation and the average/expectation values in the Tsallis-1 scheme can be written as~\cite{Parvan2015,Parvan2017a,Parvan19}, 

\bea
\label{t1prob}
p_{i} &=& \left[1+\frac{q-1}{q}\frac{\Lambda-E_{i}+\mu N_{i}}{T}\right]^{\frac{1}{q-1}}; ~~~~
  \mathlarger{\mathlarger{\sum}}\limits_{i} \left[1+\frac{q-1}{q}\frac{\Lambda-E_{i}+\mu N_{i}}{T}\right]^{\frac{1}{q-1}} =1; \nn\\
\nn\\
\text{and}~~~~~~~
\langle Q \rangle &=& \mathlarger{\mathlarger{\sum}}\limits_{i} Q_i \left[1+\frac{q-1}{q}\frac{\Lambda-E_{i}+\mu N_{i}}{T}\right]^{\frac{1}{q-1}}, 
\eea
where $\Lambda$ is a norm function, $\mu$ is the chemical potential, and $E_i$ and $N_i$ are the energy and the number of particles for the $i^{\text{th}}$ state. 

Using the integral representation of the gamma functions~\cite{Prato} for $q<1$, Eq.~\eqref{t1prob} can be rewritten as~\cite{Parvan19},  

\begin{eqnarray}
\label{t1probintrep}
 p_{i} &=& \frac{1}{\Gamma\left(\frac{1}{1-q}\right)} \int\limits_{0}^{\infty} t^{\frac{q}{1-q}} e^{-t\left[1+\frac{q-1}{q}\frac{\Lambda-E_{i}+\mu N_{i}}{T}\right]} dt; ~~~~~
\frac{1}{\Gamma\left(\frac{1}{1-q}\right)} \int\limits_{0}^{\infty} t^{\frac{q}{1-q}} e^{-t\left[1+\frac{q-1}{q}\frac{\Lambda-\Omega_{\text{G}}\left(\beta'\right)}{T}\right]} dt  = 1;
\nn\\
\text{and}~~~~~~~~~~~~~
\langle Q \rangle &=& \frac{1}{\Gamma\left(\frac{1}{1-q}\right)} \int\limits_{0}^{\infty} t^{\frac{q}{1-q}} e^{-t\left[1+\frac{q-1}{q}\frac{\Lambda-\Omega_{\text{G}}\left(\beta'\right)}{T}\right]} \langle Q \rangle_{\text{G}} (\beta') dt,
\end{eqnarray}
where 
\begin{eqnarray}
\label{1a}
 \Omega_{\text{G}}\left(\beta'\right) &=& -\frac{1}{\beta'} \ln Z_{\text{G}}\left(\beta'\right); ~~~~~~
Z_{\text{G}}\left(\beta'\right) = \sum\limits_{i} e^{-\beta'(E_{i}-\mu N_{i})};~~~~~~
\langle Q \rangle_{\text{G}}\left(\beta'\right) = \frac{1}{Z_{\text{G}}\left(\beta'\right)} \sum\limits_{i} Q_{i} e^{-\beta'(E_{i}-\mu N_{i})};
\nn\\
\text{and} ~~~~
\beta' &=& t(1-q)/qT. 
\end{eqnarray}

\section{Generalized transverse momentum spectra}
\label{sec3}

Transverse momentum ($p_{\text{T}}$) distributions of classical and quantum particles forming an ideal gas of volume $V$, are expressible in terms of the corresponding mean occupation numbers in the following way~\cite{Parvan19}, 
\bea
\label{t1spectra}
\frac{d^2N}{dp_{\text{T}}dy} = \frac{V}{(2\pi)^3} \int\limits_{0}^{2\pi} d\varphi \ p_{\mathrm{T}} \varepsilon_p \sum\limits_{\sigma} \langle n_{p\sigma}\rangle,
\eea
where $m_{\text{T}}=\sqrt{p_{\text{T}}^2+m^2}$, $\varepsilon_p=m_{\text{T}}\cosh y$, $y$ is rapidity related to the polar angle, and $\varphi$ is the azimuthal angle of emission of particles. For an azimuthally independent integrand in Eq.~\eqref{t1spectra}, the transverse momentum distribution can be obtained as,
\bea
\frac{d^2N}{dp_{\text{T}}dy} = \frac{V}{(2\pi)^2} p_{\mathrm{T}} m_{\mathrm{T}} \cosh y \sum\limits_{\sigma} \langle n_{p\sigma}\rangle. 
\eea 
Using Eq.~\eqref{t1probintrep}, we obtain the expression for the mean occupation numbers as given below~\cite{Parvan19}:
\begin{equation}\label{t1meannpintrep}
\langle n_{p\sigma}\rangle = \frac{1}{\Gamma\left(\frac{1}{1-q}\right)} \int\limits_{0}^{\infty} t^{\frac{q}{1-q}} e^{-t\left[1+\frac{q-1}{q}\frac{\Lambda-\Omega_{G}\left(\beta'\right)}{T}\right]} \langle n_{p\sigma}\rangle_{\text{G}} (\beta') dt, \;\;\;\;
\end{equation}
where 
\begin{equation}\label{meanoccugibbs}
\langle  n_{p\sigma} \rangle_{\text{G}} (\beta') = \frac{1}{e^{\beta'(\varepsilon_p-\mu)}+\eta} 
\end{equation}
in which $\varepsilon_p=\sqrt{p^2+m^2}$, $\eta=1$ for the Fermi-Dirac (FD) statistics, $\eta=-1$ for the Bose-Einstein (BE) statistics and $\eta=0$ for the Maxwell-Boltzmann (MB) statistics of particles.  

Expanding the normalization in Eq.~\eqref{t1probintrep} and the mean occupation numbers (Eq.~\ref{t1meannpintrep}) in an infinite series and using  Eq.~\eqref{t1spectra}, we obtain (see Ref.~\cite{Parvan19}),
\bea
\label{t1probnormseries}
&& \mathlarger{\mathlarger{\sum}}_{\ell=0}^{\infty}  \frac{1}{\ell !~\Gamma\left(\frac{1}{1-q}\right)} \int\limits_{0}^{\infty} t^{\frac{q}{1-q}} e^{-t\left[1+\frac{q-1}{q}\frac{\Lambda}{T}\right]}    \left[-\beta'\Omega_{G}\left(\beta'\right)\right]^{\ell}dt =1 
\\
\text{and} ~~~~~~
\frac{d^{2}N}{dp_{\text{T}}dy} &=& \frac{gV}{(2\pi)^{2}} p_{\text{T}}  m_{\text{T}} \cosh y  
\mathlarger{\mathlarger{\sum}}_{\ell=0}^{\infty}  \frac{1}{\ell!~\Gamma\left(\frac{1}{1-q}\right)} 
\int\limits_{0}^{\infty} t^{\frac{q}{1-q}} e^{-t\left[1+\frac{q-1}{q}\frac{\Lambda}{T}\right]}
\frac{\left[-\beta'\Omega_{G}\left(\beta'\right)\right]^{\ell}}{e^{\beta' (m_{\text{T}} \cosh y-\mu)}+\eta} dt, 
\nn\\
\label{t1spectraseries}
\eea
where $g$ is the spin degeneracy factor. 

In Eqs.~(\ref{t1meannpintrep}), (\ref{t1probnormseries}), and (\ref{t1spectraseries}), the thermodynamic potential $\Omega_{\text{G}}$ for classical and quantum gases in the Boltzmann-Gibbs statistics is given by the following expression~\cite{huang,Parvan19},
\bea
-\beta '\Omega_{\text{G}}(\beta') = \frac{gV}{(2\pi)^3} \int d^3 p ~~\mathrm{ln}
\Big[1+\eta e^{-\beta'(\varepsilon_p-\mu)}\Big]^{\frac{1}{\eta}} \nn\\
\label{thermpotgibbs}
\eea
for the $\eta$ values mentioned below Eq.~\eqref{meanoccugibbs}.

\section{Maxwell-Boltzmann spectrum}
\label{sec4}

For the Maxwell-Boltzmann particle statistics, $\Omega_{\text{G}}\left(\beta'\right)$ is given by~\cite{Parvan19}, 
\bea
\Omega^{\mathrm{MB}}_{\text{G}}(\beta') = -\frac{gV e^{\beta'\mu}}{2\pi^2\beta'^2} m^2 K_2 (\beta' m), 
\label{omegamassive}
\eea
where $K_2(z)$ is the modified Bessel's function of the second kind.
Putting \eqref{omegamassive} into \eqref{t1probnormseries} and \eqref{t1spectraseries}, we obtain~\cite{Parvan19},
\bea
\label{t1probnormseriesMB}
&& \mathlarger{\mathlarger{\sum}}\limits_{\ell=0}^{\infty} \frac{\omega^{\ell}}{\ell!} \frac{1}{\Gamma\left(\frac{1}{1-q}\right)} \int\limits_{0}^{\infty} t^{\frac{q}{1-q}-\ell} e^{-t\left[1+\frac{q-1}{q}\frac{\Lambda+\mu \ell}{T}\right]} \left[K_{2}\left(\beta'm\right)\right]^{\ell} dt 
=
\mathlarger{\mathlarger{\sum}}\limits_{\ell=0}^{\infty} \Phi(\ell)= 1, \\
\text{and}~~
\label{t1spectraseriesMB}
&&\frac{d^{2}N}{dp_{\text{T}}dy} = \frac{gV}{(2\pi)^{2}} p_{\text{T}}  m_{\text{T}} \cosh y
 \mathlarger{\mathlarger{\sum}}\limits_{\ell=0}^{\infty} \frac{\omega^{\ell}}{\ell!} \frac{1}{\Gamma\left(\frac{1}{1-q}\right)} 
 \int\limits_{0}^{\infty} t^{\frac{q}{1-q}-\ell} e^{-t\left[1+\frac{q-1}{q}\frac{\Lambda-m_{\text{T}} \cosh y+\mu (\ell+1)}{T}\right]} \left[K_{2}\left(\beta'm\right)\right]^{\ell} dt \nn\\
 \mathrm{for} \quad  &&\omega = \frac{gVTm^{2}}{2\pi^{2}} \frac{q}{1-q}. 
\eea
Once we have the generalized form of the classical Tsallis transverse momentum spectrum in terms of an infinite summation, 
we proceed to calculate the terms appearing in this expression.

\subsection{Zeroth order approximation or truncation at $\ell=0$}
The zeroth order contribution in the MB transverse momentum distribution is given by~\cite{Parvan19}, 
\bea
\frac{d^{2}N^{(0)}}{dp_{\text{T}}dy} &=& \frac{gV}{(2\pi)^{2}} p_{\text{T}}  m_{\text{T}} \cosh y 
\frac{1}{\Gamma\left(\frac{1}{1-q}\right)}
\int\limits_{0}^{\infty} t^{\frac{q}{1-q}}e^{-t\left[1+\frac{q-1}{q}\frac{\Lambda-m_{\text{T}} \cosh y+\mu}{T}\right]}  dt \nn\\
&=& \frac{gV}{(2\pi)^{2}} p_{\text{T}}  m_{\text{T}} \cosh y \left[1+\frac{q-1}{q}\frac{\Lambda-m_{\text{T}} \cosh y+\mu}{T}\right]^{\frac{1}{q-1}} \nn\\
\eea
Equating the $\ell=0$ term in Eq.~\eqref{t1probnormseriesMB} with 1 we obtain $\Lambda=0$, and hence the corresponding spectrum is given by~\cite{Parvan19},
\bea
\frac{d^{2}N^{(0)}}{dp_{\text{T}}dy} = \frac{gV}{(2\pi)^{2}} p_{\text{T}}  m_{\text{T}} \cosh y \left[1+\frac{1-q}{q}\frac{\varepsilon_p-\mu}{T}\right]^{\frac{1}{q-1}}.  \nn\\
\eea
As already discussed in Refs.~\cite{Parvan2017a,Parvan19}, the above expression can immediately be identified to be identical with the Tsallis-like function~\cite{TsallisTaylor,cleymansworku} widely used in literature once we replace $q \to q^{-1}$.

\subsection{First order approximation or truncation at $\ell=1$}
The first order contribution in the MB transverse momentum distribution is given by,
\bea
\frac{d^{2}N^{(1)}}{dp_{\text{T}}dy} = \frac{g V}{(2\pi)^{2}} p_{\text{T}}  m_{\text{T}} \cosh y \langle n_p^{(1)} \rangle,
\eea
where
\bea
\label{t1np1st}
\langle n_p^{(1)} \rangle = \int_0^{\infty} \frac{ \omega t^{\frac{q}{1-q}-1} } {\Gamma\left(\frac{1}{1-q}\right)}
e^{-t\left[1+\frac{q-1}{q}\frac{\Lambda-\varepsilon_p+2\mu}{T}\right]} K_{2}\left(\beta'm\right) dt \nn\\
\eea
is the first order mean occupation number. To evaluate this quantity we use the following contour-integral representation 
of $K_2$ given by \cite{pariskaminski},
\bea
K_2(\beta' m) = \intecontour  ~~\Gamma(s) \Gamma(s-2) \left(\frac{\beta' m }{2}\right)^{2-2s};~~~ 
\text{for} \quad \gamma>2  \quad \text{and} \quad \arg(\beta' m) < \pi . \nn\\
\label{K2rep}
\eea
Using the above representation and swapping the contour integration with the $t$-integration in Eq.~\eqref{t1np1st},
we get the following expression:
\bea
&&\langle n_p^{(1)} \rangle = \frac{\omega}{\Gamma\left(\frac{1}{1-q}\right)}
\intecontour \Gamma(s) \Gamma(s-2) 
 \left[\frac{(1-q) m}{2qT}\right]^{2-2s} 
 \intet ~t^{\frac{1}{1-q} - 2s }~ e^{-t\left[1+\frac{q-1}{q}\frac{\Lambda-\varepsilon_p+2\mu}{T}\right]} \nn\\
&&= \frac{\omega(1-q)^2m^2\left(1+\frac{q-1}{q}\frac{\Lambda-\varepsilon_p+2\mu}{T} \right)^{\frac{2-q}{q-1}}}{4q^2T^2\Gamma\left(\frac{1}{1-q}\right)}   
\intecontour \Gamma(s) \Gamma(s-2) \Gamma\left(\frac{2-q}{1-q}-2s\right)
\left[\frac{2qT\left(1+\frac{q-1}{q}\frac{\Lambda-\varepsilon_p+2\mu}{T}\right)}{(1-q)m} \right]^{2s}. \nn\\
\label{nav1contour}
\eea

\begin{figure}[!htb]
\begin{center}
\includegraphics[width=0.4\textwidth]{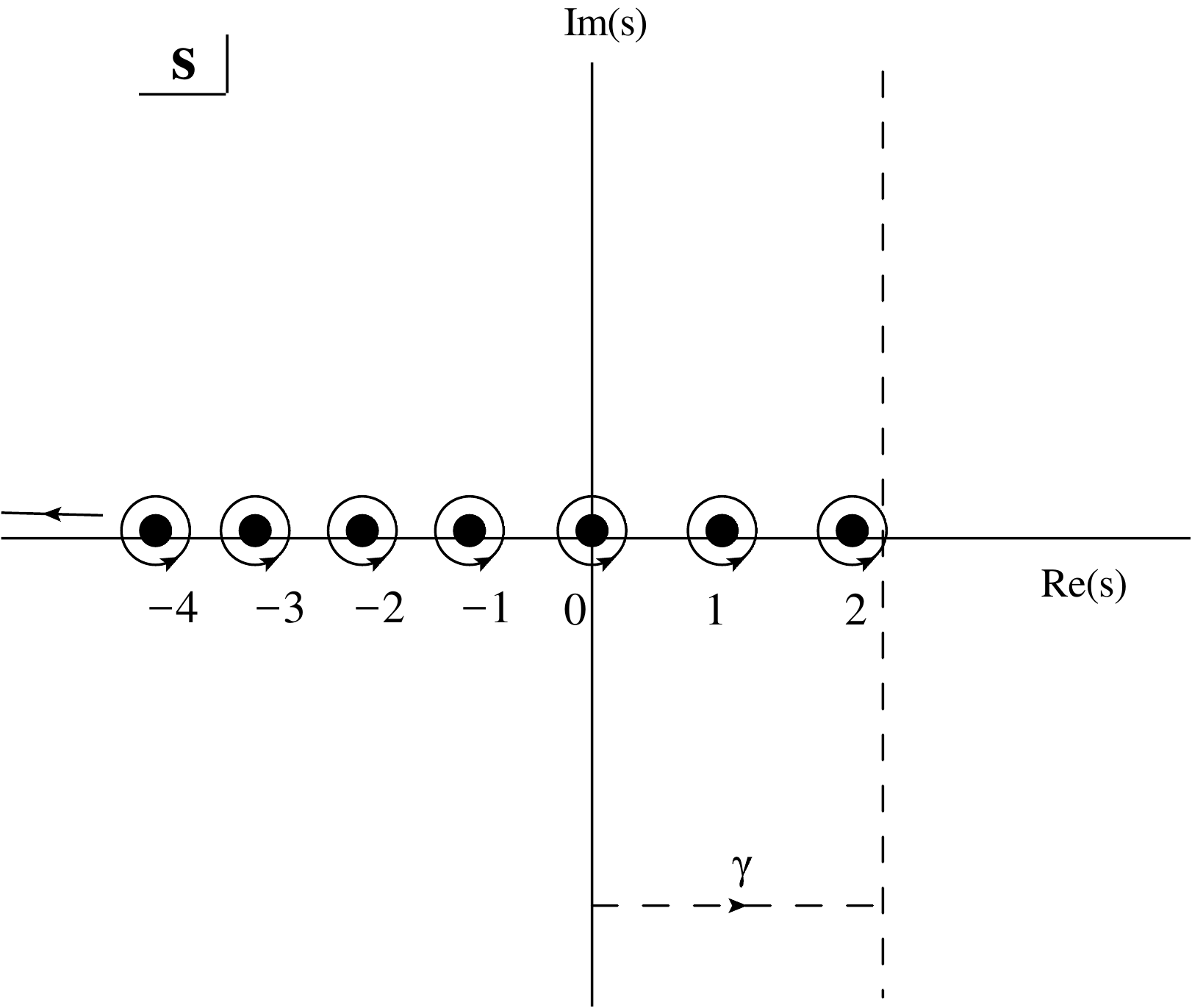}
\caption{Integration contour in Eqs.~\eqref{nav1contour} and \eqref{4a}.}
\label{nav1contourfig}
\end{center}
\end{figure}

Now, we wrap the contour anti-clockwise (see Fig.~\ref{nav1contourfig}) so that it includes the poles at $s=2,1,0,-1,-2,...,-k$ and according to the Cauchy's integral theorem, the sum of the residues at the poles multiplied by $2\pi i$ is the result of the integration. However, the condition which yields a non-divergent result of the contour integration is,
\be
\frac{2qT}{(1-q)m} \left(1+\frac{q-1}{q}\frac{\Lambda-\varepsilon_p+2\mu}{T}\right) > 1. 
\ee
\\
\\

Now, we proceed to calculate the residues. 

\begin{itemize}
\item Residue of Eq.~\eqref{nav1contour} at $s=2$ is given by,
\begin{eqnarray}
\mathcal{R}_{(2)}^{(1)} &=& \frac{\omega}{\pi i \Gamma\left(\frac{1}{1-q}\right)}
\left[\frac{qT}{(1-q)m}\right]^2  \Gamma\left(\frac{3q-2}{1-q}\right)  
 \left(1+\frac{q-1}{q}\frac{\Lambda-\varepsilon_p+2\mu}{T}\right)^{\frac{3q-2}{q-1}}.
 \label{TMB1res2}
\end{eqnarray}
\item Residue of Eq.~\eqref{nav1contour} at $s=1$ is given by,
\bea
\mathcal{R}_{(1)}^{(1)} &=& \frac{\omega (q-1)}{4 \pi i q}
\left(1+\frac{q-1}{q}\frac{\Lambda-\varepsilon_p+2\mu}{T}\right)^{\frac{q}{q-1}}.
 \label{TMB1res1}
\eea
\item Residues of Eq.~\eqref{nav1contour} at $s=-k,~k \in \mathbb{Z}^{\geq}$ can be calculated as follows,
\bea
&& \mathcal{R}_{(-k)}^{(1)} = \frac{\omega}{4\pi i} \frac{\Gamma\left(\frac{2-q}{1-q}+2k\right)}{\Gamma\left(\modz\right)k!(2+k)!}
\left[\frac{(1-q)m}{2qT}\right]^{2+2k}
\left(1+\frac{q-1}{q}\frac{\Lambda-\varepsilon_p+2\mu}{T}\right)^{\frac{2-q}{q-1}-2k}
\nn\\
&&\times \left[2 \ln \left(\frac{2qT}{m-qm}-\frac{2\Lambda-2\varepsilon_p+4\mu}{mT}\right)
-2\psi^{(0)}\left(\frac{2-q}{1-q}+2k\right) +\psi^{(0)}(k+1)+\psi^{(0)}(k+3) \right]. 
\nn\\
 \label{TMB1resminusk}
\eea
\end{itemize}
We observe that the residues at $s=-k$ contain the digamma function $\psi^{(0)}$ which is the poly-gamma 
function $\psi^{(m)}$ \cite{Bateman} at the zeroth order. Using Eqs.~\eqref{TMB1res2}, \eqref{TMB1res1} and \eqref{TMB1resminusk},  
$\langle n_p^{(1)}\rangle$ is given by,
\bea
\langle n_p^{(1)}\rangle = 2\pi i \sum_{n=-2}^{\infty} \mathcal{R}_{(-n)}^{(1)}.
\eea
Hence, the first order contribution in the Tsallis MB transverse momentum distribution can be written as,
\bea
\frac{d^{2}N^{(1)}}{dp_{\text{T}}dy}=  \frac{g V }{2\pi} p_{\text{T}}  m_{\text{T}}\cosh y  \sum_{n=-2}^{\infty} i \mathcal{R}_{(-n)}^{(1)}. 
\label{TMB1}
\eea
Eq.~\eqref{TMB1} represents the first main result of the paper. The same equation may also be obtained using the series expansion of the 
Bessel's function $K_2(z)$.

\subsection{{Second order approximation or truncation at $\ell=2$}}
The second order contribution in the MB transverse momentum distribution is given by,
\bea
\frac{d^{2}N^{(2)}}{dp_{\text{T}}dy} = \frac{gV}{(2\pi)^{2}} p_{\text{T}}  m_{\text{T}} \cosh y \langle n_p^{(2)} \rangle,
\eea
where
\bea
\label{t1np2nd}
\langle n_p^{(2)} \rangle = \frac{\omega^2}{2!} \int_0^{\infty} \frac{ t^{\frac{q}{1-q}-2} } {\Gamma\left(\frac{1}{1-q}\right)}
e^{-t\left[1+\frac{q-1}{q}\frac{\Lambda-\varepsilon_p+3\mu}{T}\right]} K^2_{2}\left(\beta'm\right) dt \nn\\
\eea
is the second order mean occupation number. To evaluate $\langle n_p^{(2)} \rangle$ we use the contour 
integral representation of $K_2^2(z)$ which is given by \cite{pariskaminski},
\begin{eqnarray}
 K_2^2(\beta' m) &=& \frac{1}{2} \intecontour \frac{\Gamma(s+2)\Gamma(s)^2 \Gamma(s-2) }{\Gamma(2s)} 
  \left(\frac{\beta' m }{2}\right)^{-2s};~~\text{for} \quad    \gamma>2, \;\;\; \arg(\beta' m) < \pi. \;\;\;\;\;\;
\label{K2sqrep}
\end{eqnarray}
Using the above representation and swapping the contour integration with the $t$-integration in
Eq.~(\ref{t1np2nd}), we get the following expression,
\bea 
\label{4a}
&& \langle n_p^{(2)} \rangle = \frac{\omega^2}{4\Gamma\left(\modz\right)} \intecontour \frac{\Gamma(s+2)\Gamma(s)^2 \Gamma(s-2) }{\Gamma(2s)}
 \left[\frac{(1-q) m }{2qT}\right]^{-2s}
 \intet ~t^{\frac{q}{1-q}-2-2s}e^{-t\left(1+\frac{q-1}{q}\frac{\Lambda-\varepsilon_p+3\mu}{T}\right)}\nn\\
&&=\frac{\omega^2\left(1+\frac{q-1}{q}\frac{\Lambda-\varepsilon_p+3\mu}{T} \right)^{\frac{2q-1}{q-1}}}{4\Gamma\left(\modz\right)} 
\intecontour \frac{\Gamma(s+2)\Gamma(s) \Gamma(s) \Gamma(s-2)\Gamma\left(\frac{2q-1}{1-q}-2s\right)}{\Gamma(2s)} 
\left[\frac{2qT\left(1+\frac{q-1}{q}\frac{\Lambda-\varepsilon_p+3\mu}{T} \right)}{(1-q)m}\right]^{2s}. \nn\\ 
\eea
Here we choose the same integration contour as that shown in Fig.~\ref{nav1contourfig}. We wrap the contour anti-clockwise
so that it includes the poles at $s = 2,1,0,-1,-2,...,-k$ and according to the Cauchy's integral theorem the sum of the residues at the poles multiplied by $2\pi i$ is the result of the integration. The condition for convergence of the integration is
\bea
\frac{2qT}{(1-q)m} \left(1+\frac{q-1}{q}\frac{\Lambda-\varepsilon_p+3\mu}{T} \right) > 1. 
\eea
Let us calculate the residues of the integration.

\begin{itemize}
\item Residue of Eq.~\eqref{4a} at $s=2$ is given by,
\begin{eqnarray}
\mathcal{R}_{(2)}^{(2)} &=& \frac{\omega^2\Gamma\left(\frac{6q-5}{1-q}\right)q^4T^4}{ \pi i \Gamma\left(\modz\right)m^4(1-q)^4} 
 \left(1+\frac{q-1}{q}\frac{\Lambda-\varepsilon_p+3\mu}{T} \right)^{\frac{6q-5}{q-1}}. \nn\\
\end{eqnarray}
\item Residue of Eq.~\eqref{4a} at $s=1$ can be written as, 
\begin{eqnarray}
 \mathcal{R}_{(1)}^{(2)} &=& - \frac{q^2 T^2 \omega ^2 \Gamma \left(\frac{3-4 q}{q-1}\right)}{2 \pi i    m^2 (q-1)^2 \Gamma \left(\frac{1}{1-q}\right)}         
 \left(1+\frac{q-1}{q}\frac{\Lambda-\varepsilon_p+3\mu}{T}\right)^{\frac{4 q-3}{q-1}}.  \nn\\
\end{eqnarray}
\item Residue of Eq.~\eqref{4a} at $s=0$ can be written as, 
\bea
\mathcal{R}_{(0)}^{(2)} &=& \frac{\omega^2\Gamma\left(\frac{2q-1}{1-q}\right)}{8 \pi i \Gamma\left(\modz\right)}
\left(1+\frac{q-1}{q}\frac{\Lambda-\varepsilon_p+3\mu}{T}\right)^{\frac{2q-1}{q-1}}
\left[\ln \left(\frac{2qT}{m-qm}-\frac{2\Lambda-2\varepsilon_p+6\mu}{m}\right)
\right. \nn\\
&& \left.
- \psi^{(0)}\left(\frac{2q-1}{1-q}\right)-\gamma_{\mathrm{E}}+\frac{5}{4} \right], \nn\\
\eea
where $\gamma_{\text{E}}= 0.57721...$ is the Euler-Mascheroni constant.
\item Residue of Eq.~\eqref{4a} at $s=-1$ can be written as,
\bea
\mathcal{R}_{(-1)}^{(2)} &=&  -\frac{\omega^2(1-q)^2m^2}{48 \pi i q^2T^2}
\left(1+\frac{q-1}{q}\frac{\Lambda-\varepsilon_p+3\mu}{T}\right)^{\frac{1}{q-1}}
\left[ \ln \left(\frac{2qT}{m-qm}-\frac{2\Lambda-2\varepsilon_p+6\mu}{m}\right)
\right.
\nn\\
&& \left.
- \psi^{(0)}\left(\modz\right)-\gamma_{\mathrm{E}}+\frac{5}{12} \right]. \nn\\
\eea
\item Residue of Eq.~\eqref{4a} at $s=-k-2,~k \in \mathbb{Z}^{\geq}$ is given by,
\bea
&& \mathcal{R}^{(2)}_{(-k-2)} =
\frac{  \omega ^2
  \Gamma (2 k+5) \Gamma
   \left(2k+\frac{2 q-3}{q-1}\right)(m-mq)^{4+2k}}
   {\pi i 4^{k+4}  (qT)^{4+2k}
    k! [(k+2)!]^2 (k+4)! \Gamma \left(\frac{1}{1-q}\right)}
    \ordertwoarg^{\frac{2 q-3}{1-q}-2k}
\nn\\
&& \times
\Bigg[\lambda _1 \bigg\{4 \psi _1+8 \psi _2
+4 \psi _3-8 \psi _4-8 \psi _5+8 \ln 
   (2)\bigg\}+\lambda _2 \bigg\{-8 \lambda _1-4 \psi _1
  -8 \psi _2-4 \psi_3+8 \psi _4+8 \psi _5-8 \ln (2)\bigg\} 
   \nn\\
   &&+ \psi _2 \bigg\{4 \psi _3-8 \psi _4-8 \psi _5+4 \ln (4)\bigg\}+\psi_3 \bigg\{-4 \psi _4-4 \psi _5+\ln (16)\bigg\}
+\psi _1 \bigg\{4 \psi _2+2\psi _3-4 \psi _4-4 \psi _5+\ln (16)\bigg\}
  \nn\\ 
&&   
+\psi _5 \bigg\{8 \psi _4-8
   \ln (2)\bigg\}
   +4 \ln ^2(2)+4 \lambda _1^2+4 \lambda
   _2^2+\psi _1^2+4 \psi _2^2
   +\psi _3^2+4 \psi _4^2+4 \psi _5^2-\psi
   _{11}-2 \psi _{21}-\psi _{31}+4 \psi _{41}
   \nn\\
   &&
   +4 \psi _{51}-8 \psi _4 \ln(2) \Bigg].
\label{kminus2}
\eea
\end{itemize}
In the above equation,
\begin{itemize} 
\item $\lambda_1$, and $\lambda_2$ are given by,
\bea
&&\ln\left(1+\frac{q-1}{q}\frac{\Lambda-\epsilon+3\mu}{T}\right) = \lambda_1;
~\ln\left(\frac{m-mq}{qT}\right) = \lambda_2. \nn\\
\eea
\item $\psi_j$, and $\psi_{j1}$ ($j=1\rightarrow5$) are given by, 
\bea
\psi^{(0)}(k+2j-1)&=& \psi_j;~\psi^{(1)}(k+2j-1) = \psi_{j1}~~ (\text{for}~j=1\rightarrow 3); \nn\\
\psi^{(0)}(2k+5) &=& \psi_4;~\psi^{(1)}(2k+5)   = \psi_{41};~ 
\psi^{(0)}\left(\frac{2q+2kq-2k-3}{q-1}\right) = \psi_5;
\nn\\
\psi^{(1)}\left(\frac{2q+2kq-2k-3}{q-1}\right) &=& \psi_{51}. \nn\\
\eea
\end{itemize}

The average occupation number at the second order is given by,
\bea
\langle n_p^{(2)}\rangle= 2\pi i 
 \sum_{n=-2}^{\infty} \mathcal{R}_{(-n)}^{(2)}. 
\eea
Hence, the second order contribution in the Tsallis MB transverse momentum distribution can be written as,
\bea
\frac{d^{2}N^{(2)}}{dp_{\text{T}}dy}=  \frac{g V }{2\pi} p_{\text{T}}  m_{\text{T}}\cosh y  \sum_{n=-2}^{\infty} i \mathcal{R}_{(-n)}^{(2)}. 
\label{TMB2}
\eea
Eq.~\eqref{TMB2} gives us the second main result of the paper.
\section{Quantum spectra: the zeroth order terms}
\label{sec5}

The zeroth order Tsallis Bose-Einstein and Fermi-Dirac spectra have also been computed in Ref.~\cite{Parvan19} in terms 
of an infinite summation. However, it is possible to express the infinite summation presented in Eq.~(82) of that paper in terms 
of the Hurwitz-zeta function $\zeta(s,a)$ \cite{Bateman} as discussed in the next two sub-sections. 
\subsection{Bose-Einstein}
The Boltzmann-Gibbs grand thermodynamic potential for the bosons obtained from Eq.~\eqref{thermpotgibbs} is given by,
\bea
\Omega_{\text{G}}^{\text{BE}}\left(\beta'\right) = -\frac{gV}{2\pi^2\beta'^2} \mathlarger{\mathlarger{\sum}}_{n=1}^{\infty} \frac{e^{n\beta'\mu}}{n^2} m^2 K_2\left(n\beta'm\right).
\eea
At the zeroth order of Eq.~\eqref{t1spectraseries}, the Tsallis bosonic spectrum is as follows,
\bea
\frac{d^{2}N^{(0)}_{\text{BE}}}{dp_{\text{T}}dy} &=& \frac{gV}{(2\pi)^{2}}  \frac{p_{\text{T}}  m_{\text{T}} \cosh y}{\Gamma\left(\frac{1}{1-q}\right)} 
\int\limits_{0}^{\infty}  
\frac{t^{\frac{q}{1-q}}e^{-t}}{e^{\beta' (m_{\text{T}} \cosh y-\mu)}-1} dt \nn\\
&=& \frac{gV}{(2\pi)^{2}}  \frac{p_{\text{T}}  m_{\text{T}} \cosh y}{\Gamma\left(\frac{1}{1-q}\right)} 
\mathlarger{\mathlarger{\sum}}_{s=1}^{\infty}
 \int\limits_{0}^{\infty}  dt ~t^{\frac{q}{1-q}}
e^{-t-s\beta'(m_{\text{T}} \cosh y-\mu)} \nn\\
&=& \frac{gV}{(2\pi)^{2}}  p_{\text{T}}  m_{\text{T}} \cosh y
\mathlarger{\mathlarger{\sum}}_{s=1}^{\infty} \left[1+\frac{s(1-q)}{qT}(m_{\text{T}} \cosh y-\mu)\right]^{-\frac{1}{1-q}} \nn\\
&=& \frac{gV}{(2\pi)^{2}}  p_{\text{T}}  m_{\text{T}} \cosh y
\left[\frac{(1-q) (m_{\text{T}} \cosh y-\mu )}{q
   T}\right]^{-\frac{1}{1-q}} 
\zeta
   \left(\frac{1}{1-q},1+\frac{q T}{(1-q)
   (m_{\text{T}} \cosh y-\mu )}\right).
   \label{TBE0}
\eea
\subsection{Fermi-Dirac}
The Boltzmann-Gibbs grand thermodynamic potential for the fermions obtained from Eq.~\eqref{thermpotgibbs} is given by,
\bea
\Omega_{\text{G}}^{\text{BE}}\left(\beta'\right) = -\frac{gV}{2\pi^2\beta'^2} \mathlarger{\mathlarger{\sum}}_{n=1}^{\infty} (-1)^{n+1}\frac{e^{n\beta'\mu}}{n^2} m^2 K_2\left(n\beta'm\right).
\nn\\
\eea
At the zeroth order of Eq.~\eqref{t1spectraseries}, the Tsallis fermionic spectrum looks like,
\bea
&& \frac{d^{2}N^{(0)}_{\text{FD}}}{dp_{\text{T}}dy} = \frac{gV}{(2\pi)^{2}}  \frac{p_{\text{T}}  m_{\text{T}} \cosh y}{\Gamma\left(\frac{1}{1-q}\right)} 
\int\limits_{0}^{\infty}  
\frac{t^{\frac{q}{1-q}}e^{-t}}{e^{\beta' (m_{\text{T}} \cosh y-\mu)}+1} dt \nn\\
&=& \frac{gV}{(2\pi)^{2}}  \frac{p_{\text{T}}  m_{\text{T}} \cosh y}{\Gamma\left(\frac{1}{1-q}\right)} 
\mathlarger{\mathlarger{\sum}}_{s=1}^{\infty} (-1)^{s-1}
 \int\limits_{0}^{\infty}  dt ~t^{\frac{q}{1-q}}
e^{-t-s\beta'(m_{\text{T}} \cosh y-\mu)} \nn\\
&=& \frac{gV}{(2\pi)^{2}}  p_{\text{T}}  m_{\text{T}} \cosh y
\mathlarger{\mathlarger{\sum}}_{s=1}^{\infty} (-1)^{s-1} \left[1+\frac{s(1-q)}{qT}(m_{\text{T}} \cosh y-\mu)\right]^{-\frac{1}{1-q}} \nn\\
&=& \frac{gV}{(2\pi)^{2}}  p_{\text{T}}  m_{\text{T}} \cosh y
\left[\frac{2(1-q) (m_{\text{T}} \cosh y-\mu )}{q
   T}\right]^{-\frac{1}{1-q}} 
\left[ \zeta
   \left(\frac{1}{1-q},\frac{1}{2}+\frac{q T}{2(1-q)
   (m_{\text{T}} \cosh y-\mu )}\right) \right. 
   \nn\\
&& 
\left. - 
   \zeta
   \left(\frac{1}{1-q},1+\frac{q T}{2(1-q)
   (m_{\text{T}} \cosh y-\mu )}\right) \right].
   \label{TFD0}
\eea
It is worth noting that unlike the classical case, the zeroth order quantum Tsallis spectra
does not resemble the phenomenological quantum Tsallis distributions when the $q\rightarrow q^{-1}$ replacement is done. 
However, it can be shown that if one takes the factorization approximation, the forms of the zeroth order quantum Tsallis 
distributions and the phenomenological spectra display some similarity. This point will be discussed in the next section.
\section{Quantum spectra: factorization approximation of the zeroth order terms}
\label{sec6}

For the factorization approximation we use the following substitution \cite{hasegawa},
\bea
\left[1+\frac{s(1-q)}{qT}(m_{\text{T}} \cosh y-\mu)\right]^{-\frac{1}{1-q}}
\approx
\left[1+\frac{(1-q)}{qT}(m_{\text{T}} \cosh y-\mu)\right]^{-\frac{s}{1-q}}. \nn\\
\label{hasegawafact}
\eea
Using Eq.~\eqref{hasegawafact}, the factorized (denoted by the subscript `F') zeroth order Tsallis Bose-Einstein distribution becomes,
\bea
\frac{d^{2}N^{(0)}_{\text{BE,F}}}{dp_{\text{T}}dy} =  \frac{gV}{(2\pi)^{2}}  
\frac{p_{\text{T}}  m_{\text{T}} \cosh y}{\left[1+\frac{(1-q)}{q T}(m_{\text{T}} \cosh y-\mu )\right]^{\frac{1}{1-q}}-1}. \nn\\
\label{TBE0F}
\eea
Similarly, the factorized zeroth order Tsallis Fermi-Dirac distribution is given by,
\bea
\frac{d^{2}N^{(0)}_{\text{FD,F}}}{dp_{\text{T}}dy} = \frac{gV}{(2\pi)^{2}}  
\frac{p_{\text{T}}  m_{\text{T}} \cosh y} {\left[1+\frac{(1-q)}{q T}(m_{\text{T}} \cosh y-\mu )\right]^{\frac{1}{1-q}}+1}. \nn\\
\label{TFD0F}
\eea
Replacing $q$ with $q^{-1}$ in Eqs.~\eqref{TBE0F}, and \eqref{TFD0F}, we obtain the distribution functions 
\begin{eqnarray}
\left. \frac{d^{2}N^{(0)}_{\text{BE/FD,F}}} {dp_{\text{T}}dy} \right|_{q\rightarrow q^{-1}}=
\frac{gV}{(2\pi)^{2}}  
\frac{p_{\text{T}}  m_{\text{T}} \cosh y}
{\big[1+\frac{(q-1)}{T}(m_{\text{T}} \cosh y-\mu) \big]^{\frac{q}{q-1}}\mp1},
\label{tsallisqua}
\end{eqnarray}
which are similar (but not exactly equal) to the quantum Tsallis-like distributions proposed in \cite{millerTsFD,TsFDPLA} 
\begin{eqnarray}
\frac{d^{2}N_{\text{BE/FD}}} {dp_{\text{T}}dy} =
\frac{gV}{(2\pi)^{2}}  
\frac{p_{\text{T}}  m_{\text{T}} \cosh y}
{\big[1+\frac{(q-1)}{T}(m_{\text{T}} \cosh y-\mu) \big]^{\frac{1}{q-1}}\mp1}. 
\label{tsallisquapheno}
\end{eqnarray}

\section{Results}
\label{sec7}

\begin{figure*}[!htb]
\vspace*{+1cm}
\minipage{0.45\textwidth}
\includegraphics[width=\linewidth]{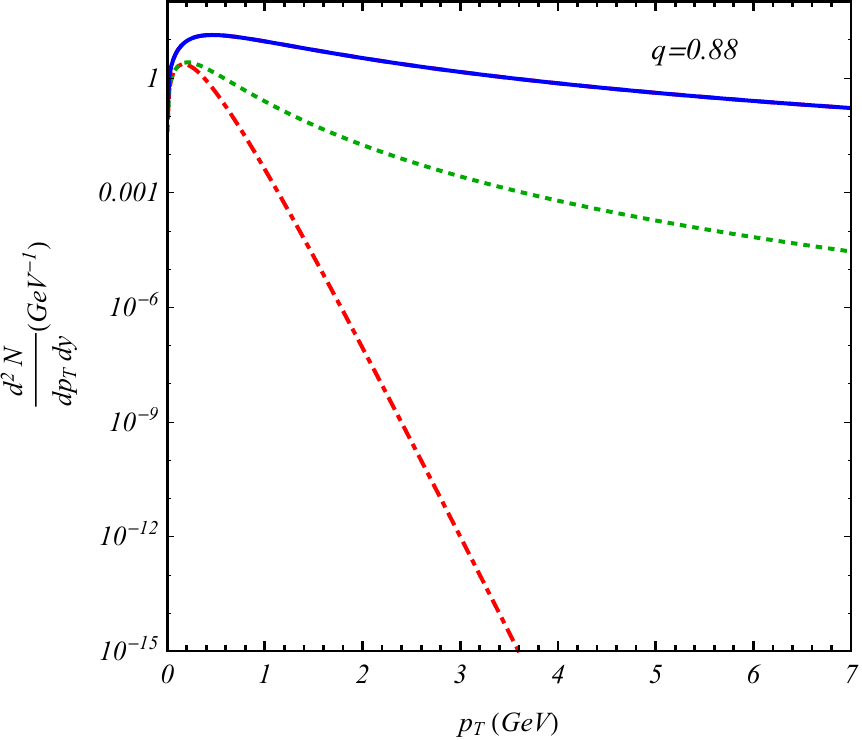}
\endminipage\hfill
\minipage{0.45\textwidth}
\vspace*{-0cm}
\hspace*{-0cm}
\includegraphics[width=\linewidth]{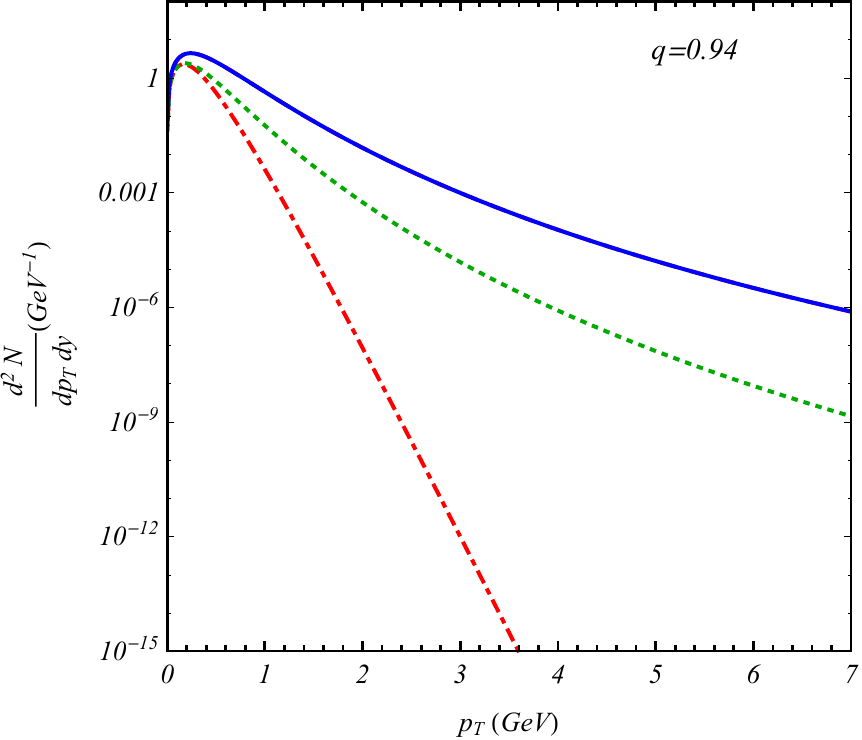}
\endminipage\hfill
\minipage{0.45\textwidth}
\vspace*{+1cm}
\hspace*{-0.1cm}
\includegraphics[width=\linewidth]{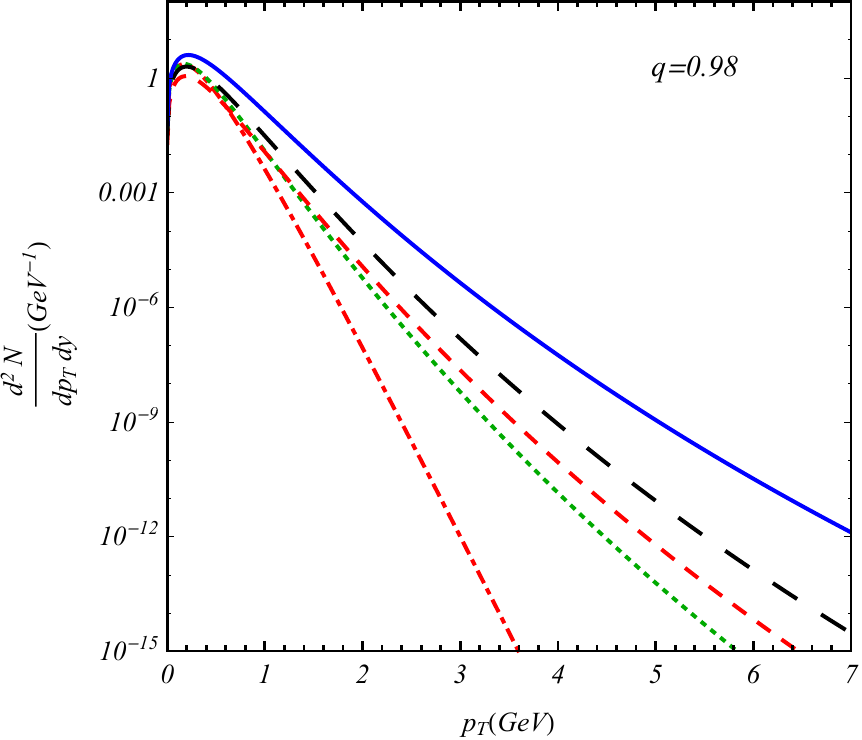}
\endminipage\hfill
\minipage{0.45\textwidth}
\vspace*{+1cm}
\hspace*{-0cm}
\includegraphics[width=\linewidth]{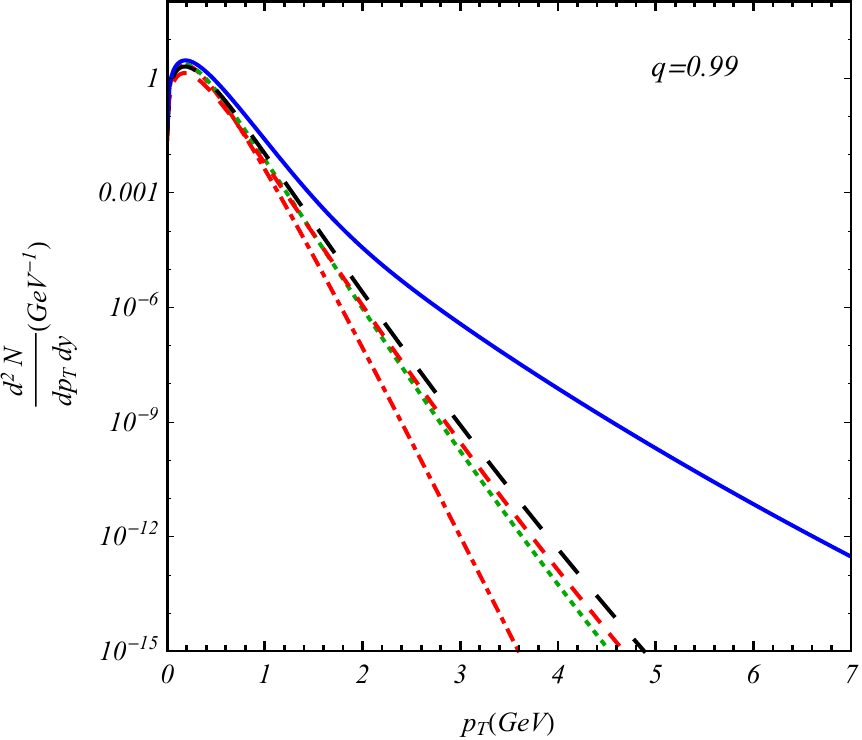}
\endminipage\hfill
\caption{Spectra of the Maxwell-Boltzmann massive particles in the Tsallis-1 statistics at mid-rapidity ($y=0$) for different values of the entropic parameter $q$ when temperature $T=82$ MeV, chemical potential $\mu=0$, radius $R=4$ fm and mass $m=139.57$ MeV (pion mass). The solid(blue), dotted(green) and dot-dashed(red) lines correspond to the exact Tsallis-1 statistics, zeroth-term approximation and the Boltzmann-Gibbs statistics $(q=1)$, respectively. The red
short-dashed (black long-dashed) lines corresponds to analytical calculation up to the first (second) order. For $q=0.88$ and $q=0.94$, the analytical first order results overlap with the numerically obtained unapproximated results.}
\label{fig2}
\end{figure*}

\begin{figure}[!htb]
\begin{center}
\includegraphics[width=0.45\textwidth]{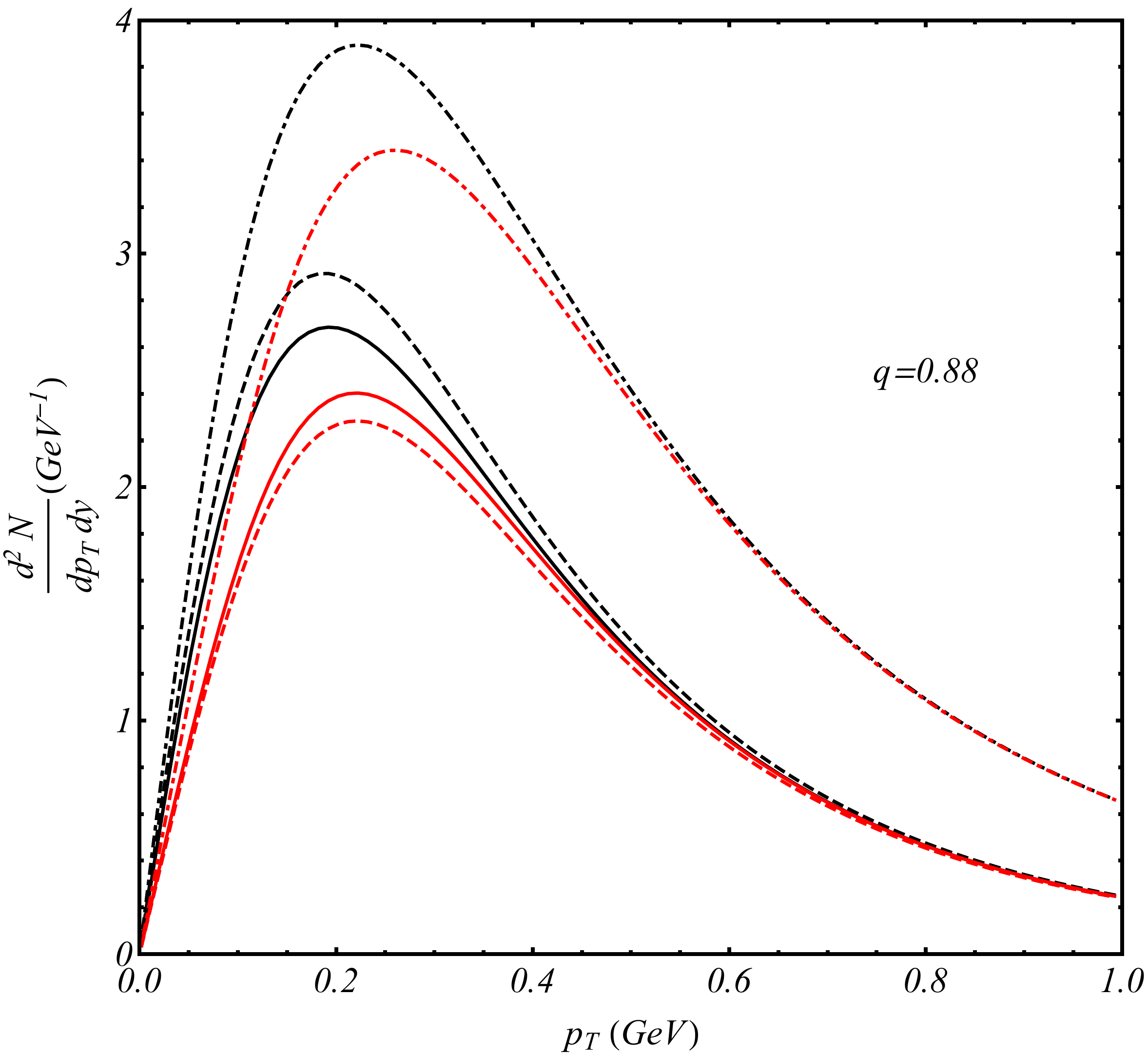}
\caption{Spectra of the Bose-Einstein (black) and Fermi-Dirac (red) massive particles in the Tsallis-1 statistics at mid-rapidity ($y=0$) for 
$q=0.88$ when temperature $T=82$ MeV, chemical potential $\mu=0$, radius $R=4$ fm and mass $m=139.57$ MeV (pion mass). The dot-dashed curves denote the quantum Tsallis phenomenological distributions, the dashed lines are the zeroth order terms, and the solid lines 
denote the factorized zeroth order distributions.}
\label{BEFD0fact}
\end{center}
\end{figure}

In Fig.~\ref{fig2}, we plot the exact hadronic spectra  (see Eq.~\ref{t1spectraseriesMB}) for four values of the entropic parameter $q =0.88, 0.94, 0.98$, and 0.99 for a particle with the pion mass $m=139.570$ MeV. Temperature $T=82$ MeV, chemical potential $\mu=0$ MeV, and radius $R=4$ fm. We compare the exact spectra with the zeroth, first, and second order approximated spectra as well as with the Boltzmann-Gibbs counterpart. 

As seen in Ref.~\cite{Parvan19}, the exact Tsallis transverse momentum spectra diverge if one includes all the terms in the series representing the spectra in Eq.~(\ref{t1spectraseriesMB}). A regularization scheme which cuts-off the series at a value of $\ell=\ell_0$ corresponding to the minimum of the probability function $\phi(\ell)$ in Eq.~(\ref{t1probnormseriesMB}) and the normalization of the probabilities yields the numerical value of the potential $\Lambda$. For the zeroth, first, and second order approximated calculations, $\Lambda$ is calculated by truncating the series at $\ell_0 =0, 1$ and 2 respectively and by solving the normalization equation.

It is observed in Fig.~\ref{fig2} that for $q=0.88,~0.94$ and for the given values of mass, chemical potential and temperature, the first order spectra overlap with the exact calculations. For $q=0.98$, and 0.99, however, the first and the second order truncation are not good approximations and the corresponding spectra deviate largely from the exact calculations. 

In Fig.~\ref{BEFD0fact} we compare the zeroth order approximated quantum Tsallis distributions, their factorized counterparts, and the phenomenological quantum Tsallis distributions for $q=0.88$, $T=82$ MeV, $\mu=0$ MeV, and radius $R=4$ fm at the mid-rapidity $y=0$. It is observed that at low $p_{\text{T}}$ the zeroth order distributions and their factorized counterparts differ upto 10-20 \%, but this difference diminishes for higher $p_T$ values. 
However, this difference is much more prominent between the former ones and the phenomenological distributions, and this difference increases with
$p_{\text{T}}$. 

\section{Summary, conclusions, and outlook}
\label{sec8}

Following is a summary of the results we have obtained in this paper,
\\
\begin{itemize}
\item Analytical results for the first, and the second order corrections to the widely
used Tsallis-like classical transverse momentum distribution have been obtained (Eqs.~\ref{TMB1} and \ref{TMB2}). 
\\
\item Analytical closed form of the zeroth order terms in the quantum Tsallis spectra has 
been calculated (Eqs.~\ref{TBE0} and \ref{TFD0}). 
\\
\item Unlike the classical case, the quantum zeroth order terms are similar (but not exactly equal) to 
the phenomenological distributions only under the factorization approximation when the $q\rightarrow q^{-1}$ 
substitution is done (Eqs.~\ref{tsallisqua} and \ref{tsallisquapheno}). 
\end{itemize}

Tsallis transverse momentum spectra beyond the zeroth order approximation may be important in certain scenarios 
and hence, analytical formulae instead of a numerical integration to find out the higher order contributions will render 
data analysis procedure much easier. Also, from a more general perspective, the mathematical set-up presented in this
paper may help obtain analytical closed form results in many other cases where the Bessel's functions appear. It is
noteworthy that in Eqs.~\eqref{TMB1}, and \eqref{TMB2}, the distributions are expressed in terms of an infinite summation
owing to the infinite number of poles at Re$(s) < 0$. However, it has been explicitly verified that the residues at those poles
have drastically diminishing contributions, and for all the practical purposes, considering only a few terms of Eqs.~\eqref{TMB1resminusk},
and \eqref{kminus2} suffices.

In the experimental and the phenomenological studies so far, the zeroth order term in the Tsallis Maxwell-Boltzmann spectrum has 
been used. When we calculate the zeroth order term in the quantum statistics, no resemblance with the quantum Tsallis phenomenological 
distributions used in the literature could be established.  Even the fact that the factorization approximation of Eqs.~\eqref{TBE0}, and \eqref{TFD0} failed to show their congruence with the phenomenological distributions strengthens this argument. Although it has been shown in Ref.~\cite{TsFDPLA2} that using the factorization 
approximation in the Tsallis-2 (another scheme) mean occupation number right from the beginning leads to the phenomenological quantum distributions, this calculation contains
an inconsistent definition of the average values, and the correct form of the distribution functions are computed in Ref.~\cite{Parvan:2019hqf}.
Also, from another recent work \cite{tsallistft} it is observed that the single particle distributions appearing in the Tsallis two-point functions are the factorized zeroth order quantum distributions. Hence, we propose the usage of the analytical closed forms of the quantum Tsallis distributions calculated in Eqs.~\eqref{TBE0}, and \eqref{TFD0} as they are derivable from a fundamental theory like statistical mechanics. It will also be interesting to extend the quantum calculations beyond the zeroth order.

The results obtained in this paper may have several important implications. One of them may be the modification of the analytic expression of the 
Tsallis classical \cite{bcmprd} and quantum thermodynamic variables which will influence the non-extensive equations of state \cite{TsallisMIT1,
TsallisMIT2}. It may be possible that for dense systems, beyond the zeroth order terms become important. One more notable application will be to 
revisit the non-extensive behaviour of the QCD strong coupling studied in \cite{Javidan:2020lup}. This paper uses the expression of the non-extensive QCD coupling derived in \cite{sukanyatsallis} using the phenomenological quantum Tsallis distributions and successfully treats the deviation between the theoretical and experimental results at low energy. However, it will be interesting to see how the present results affect their finding.

\subsection*{Acknowledgement}
Authors acknowledge the support from the joint project between the JINR and IFIN-HH.


%

\end{document}